\documentclass{jltp} 
\usepackage{graphicx}
\usepackage{amssymb}
\usepackage{amsmath}
\usepackage{amsfonts}
\usepackage{bm}
\usepackage{graphicx}
\bmdefine\bomega{\omega} \bmdefine\bOmega{\Omega}
\bmdefine\bnabla{\nabla} \bmdefine\bkappa{\kappa}
\bmdefine\bphi{\phi}
\title{Dynamic Remanent Vortices in Superfluid $^3$He-B}
\date{today}
\author{R.E. Solntsev$^{\dag}$, R. de Graaf $^{\dag}$, V.B.~Eltsov$^{\dag \ddag}$,
R.~H\"anninen$^{\dag}$, \\ and M.~Krusius$^{\dag}$}
\address{$^\dag$ Low Temperature Laboratory, Helsinki University of Technology,
Finland \\ $^\ddag$ Kapitza Institute for Physical Problems,
Kosygina 2, 119334 Moscow,  Russia} \runninghead{\bfseries R.E.
Solntsev \textit{et al.}}{Dynamic remanent vortices in $^3$He-B}

\begin{document}
\maketitle \vspace{-8mm}

\begin{abstract}
We investigate the decay of vortices in a rotating cylindrical
sample of $^3$He-B, after rotation has been stopped. With
decreasing temperature vortex annihilation slows down as the
damping in vortex motion, the mutual friction dissipation $\alpha
(T)$, decreases almost exponentially. Remanent vortices then
survive for increasingly long periods,  while they move towards
annihilation in zero applied flow. After a waiting period $\Delta
t$  at zero flow, rotation is reapplied and the remnants evolve to
rectilinear vortices. By counting these lines, we measure at
temperatures above the transition to turbulence $\sim 0.6 \,
T_{\rm c}$ the number of remnants as a function of $\alpha (T)$
and $\Delta t$. At temperatures below the transition to turbulence
$T \lesssim 0.55 \, T_{\rm c}$, remnants expanding in applied flow
become unstable and generate in a turbulent burst the equilibrium
number of vortices. Here we measure the onset temperature $T_{\rm
on}$ of turbulence as a function of $\Delta t$, applied flow
velocity $\bm{v} = \bm{v}_{\rm n} - \bm{v}_{\rm s}$, and length of
sample $L$.


\end{abstract}

\vspace{-5mm}
In superfluid $^3$He-B the length scale of the vortex core radius
is the superfluid coherence length $\xi (T) = \xi (0)\,
(1-T/T_{\rm c})^{-1/2}$, with $\xi (0) = 12$ --- 65\,nm for
pressures $P=34$ --- 0\,bar. Apparently this is large enough
compared to the typical surface roughness of our fused quartz
sample cylinder that it has not been possible to measure surface
pinning or surface friction. A sensitive test is to check for
vortex formation when flow is applied, to find out whether
vortices have remained pinned at surface traps for indefinitely
long waiting periods $\Delta t$ in zero applied flow. If $\Delta
t$ is long enough, all remnants have annihilated, unless they
become permanently pinned. At mK temperatures the energy barriers
associated with such pinning traps are typically large compared to
thermal energies and once loaded, a trap remains occupied as long
as the flow velocity $v$ is kept at sufficiently low level. To
release a remnant from the trap, the required applied flow
velocity is $v_{\rm c} \approx 0.8 \kappa/(2\pi r_{\rm o})$, if
the remnant is modeled as a half circle of radius $r_{\rm o}$
attached at both ends to the cylindrical
wall.\cite{NucleationJLTP}

\begin{figure}[t]
\begin{center}
\includegraphics[width=0.72\linewidth]{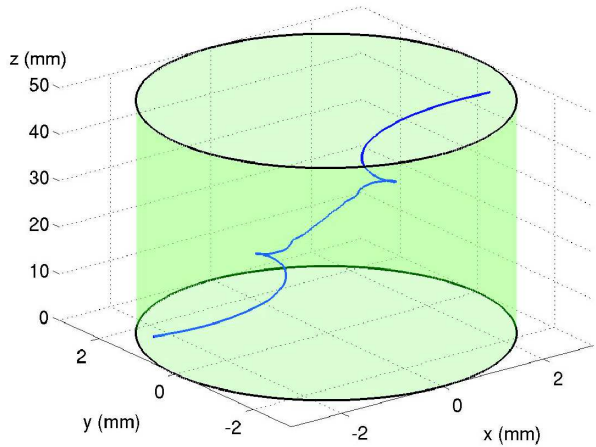}
\vspace{-3mm} \caption{A single remanent vortex in an ideal
cylinder (with no pinning or surface friction) at zero applied
flow in $^3$He-B ($0.40\,T_{\rm c}$, 29\,bar). Both ends of the
vortex move slowly in spiral motion around the cylindrical wall in
the image flow field created by the vortex itself. This snapshot
shows the configuration after 1.3\,h of motion towards final
annihilation which takes 2.3\,h. The vortex was initially straight
but tilted, with its ends at (-1,0,0) and (1,0,50) on the two flat
end plates of the cylinder, which is 50\,mm long and 3\,mm in
radius.  } \label{RemnantVortex}
\end{center}
\vspace{-8mm}
\end{figure}

This measurement is straightforward, but it involves three
complications: (i) Remnants can exist on different length scales.
The length scale determines the flow velocity at which a
particular remnant is inflated out of the pinning trap. (ii) In
the presence of only a few isolated pinning traps remanence has
stochastic nature since during random annihilation a particular
pinning site may or may not become loaded with a remnant. (iii)
The dynamical behavior of vortices changes with temperature
because of the strong temperature dependence of mutual friction
dissipation $\alpha (T,P)$.\cite{bevan} At high temperatures
vortex motion is rapid and number conserving. When rotation is
here suddenly stopped from an equilibrium vortex state at $\Omega
(t=0)$, the decay of rectilinear vortices is of the
form\cite{Dynamics} $N(t) \propto (1+ t/\tau)^{-1}$, with a decay
time $\tau = [2 \alpha \Omega (0)]^{-1}$.

In contrast, at low temperatures the motion of an isolated vortex
at zero applied flow becomes sluggish, as known from $^4$He-II.
Below $0.5\,T_{\rm c}$ the motion of the last vortex towards total
annihilation may require hours, as seen in the numerically
calculated\cite{simulation} example in Fig.~\ref{RemnantVortex}.
In this situation it becomes experimentally difficult to
distinguish between {\it pinned remnants} and {\it dynamic
remnants}.

\textbf{Experimental setup:} We measure with non-invasive NMR
techniques the number of vortex lines at 29.0\,bar pressure in a
sample cylinder with radius $R=3\,$mm and length $L=110\,$mm. The
axis of the cylinder is tilted by an angle of $0.64^{\circ}$ from
the rotation axis. NMR detection coils are located close to both
ends of the sample.\cite{ExpSetup} The sample can also be studied
in a second configuration: By sweeping up a magnetic barrier field
$H_{\rm b} > H_{\rm AB} (T,P)$ over the central section of the
long sample, a layer of $^3$He-A is created. It divides the
cylinder in two separated $^3$He-B sections which can be studied
independently, because the A phase layer acts as a
barrier\cite{ExpSetup} for vortices. With a fixed current ($I_{\rm
b} = 6.0\,$A) in the barrier magnet, the B-phase sections have the
lengths $L_{\rm t}=44\,$mm (41\,mm) for the top and $L_{\rm
b}=54\,$mm (51\,mm) for the bottom sections at $0.57\,T_{\rm c}$
($0.70\, T_{\rm c}$). By rotating the sample cylinder with angular
velocity $\Omega$, the maximum flow velocity is reached at the
cylindrical wall, $v(\Omega,N) = \Omega R - \kappa N/(2\pi R)$,
where $N$ is the number of rectilinear vortex lines in a central
vortex cluster. $N$ is determined by comparing the NMR line shape
to one measured with a known number of vortices (taking into
account the tilt angle between axes).\cite{NucleationJLTP}

\begin{figure}[t]
\begin{center}
\centerline{\includegraphics[width=0.93\linewidth]{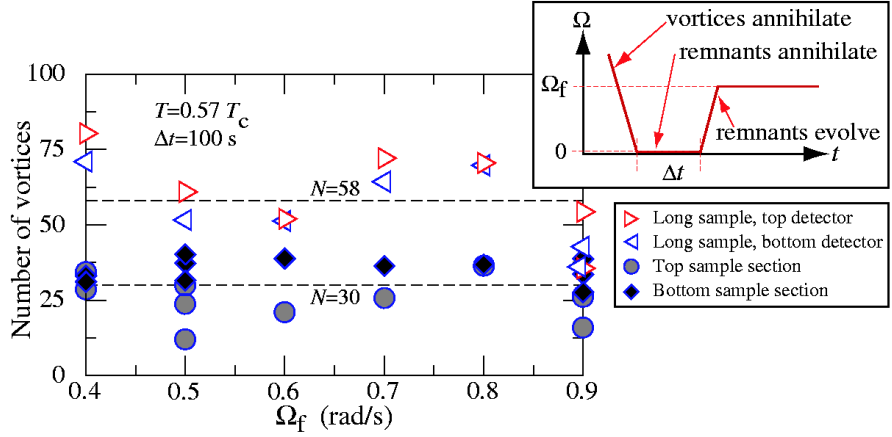}}
\vspace{-2mm} \caption{Number of remnants $\cal{N}$ after a
waiting period $\Delta t = (100 \pm 5)\,$s, measured as a function
of $\Omega_{\rm f}$.  The average for the A-phase separated top
and bottom sections is 30 and for the long sample 58. {\em
(Insert)} Measuring procedure and rotation drive.}
\label{RemnantVorNumber}
\end{center}
\vspace{-10mm}
\end{figure}

\textbf{Measuring procedure:} The measurement is started from a
state with large $N$ at $\Omega \gtrsim 1.7\,$rad/s which is
decelerated at 0.02\,rad/s$^2$ to zero rotation and kept there for
a waiting period $\Delta t$. Rotation is then increased (in the
same direction) at 0.01\,rad/s$^2$ to $\Omega_{\rm f}$ where it is
kept constant (see insert in Fig.~\ref{RemnantVorNumber}). The NMR
line shape measured at $\Omega_{\rm f}$, after all transients have
decayed, is compared to a calibration with a known number of
rectilinear vortices.\cite{ExpSetup} This gives $N$ in the final
state at $\Omega_{\rm f}$.  At temperatures above the transition
to turbulence $T_{\rm on}$ we equate $N$ with the number of
remnants $\cal{N}$ at the end of the waiting period $\Delta t$,
while below $T_{\rm on}$ the equilibrium vortex state is
created.\cite{SlowPrecursor}  In both cases a remnant which starts
to evolve has to be above a minimum size which at $\Omega_{\rm f}
= 0.9\,$rad/s is $r_{\rm o} \gtrsim 3\,\mu$m. In comparison, the
measured roughness of the quartz surface is typically $\lesssim
1\,\mu$m. The rate of rotation increase $ d\Omega / dt $ to the
final state $\Omega_{\rm f}$ was found not to influence the
results in our operating range $d\Omega / dt \sim 0.001$ ---
0.04\,rad/s$^2$.

\begin{figure}[t]
\begin{center}
\includegraphics[width=0.48\linewidth]{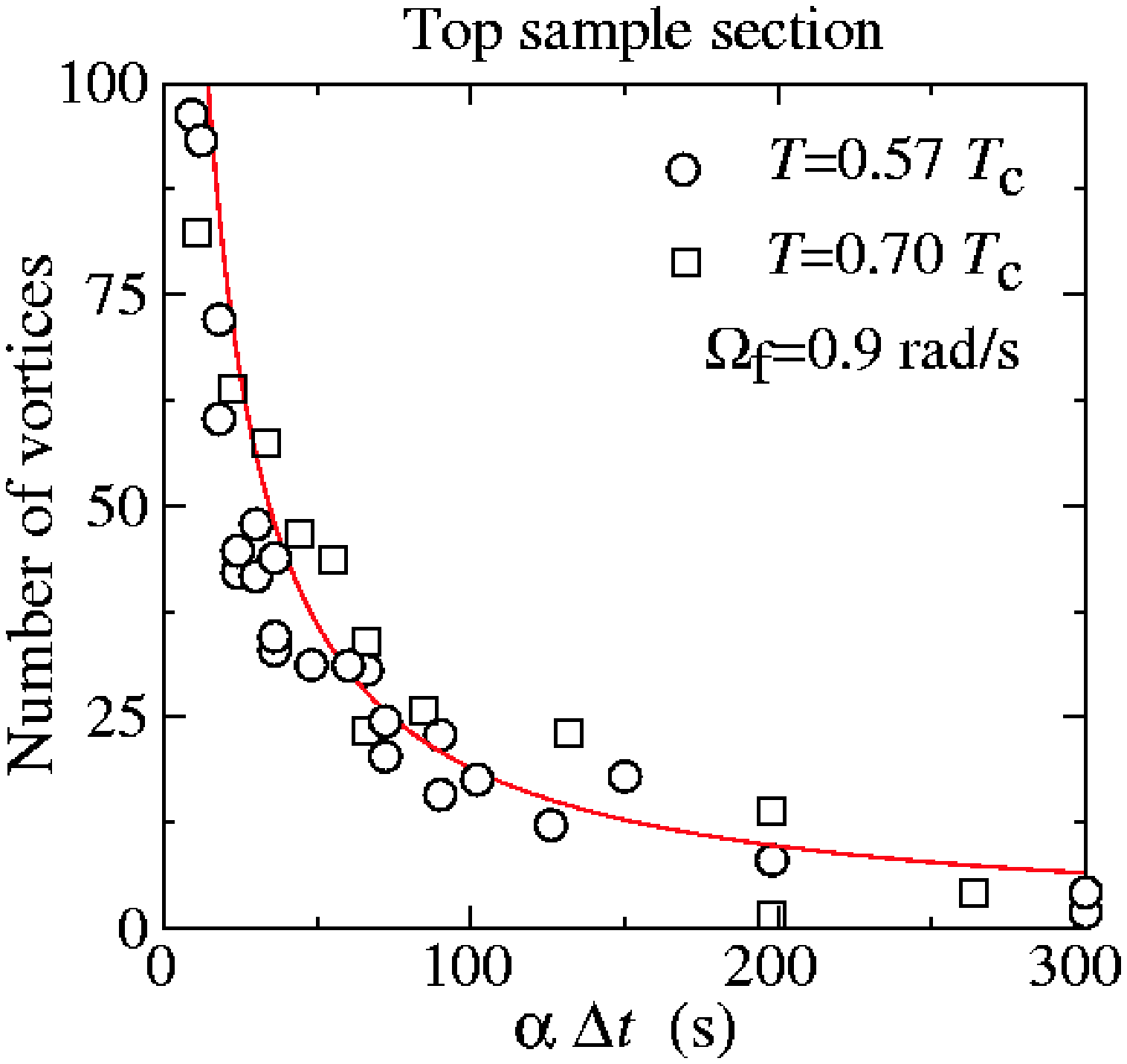}
\includegraphics[width=0.48\linewidth]{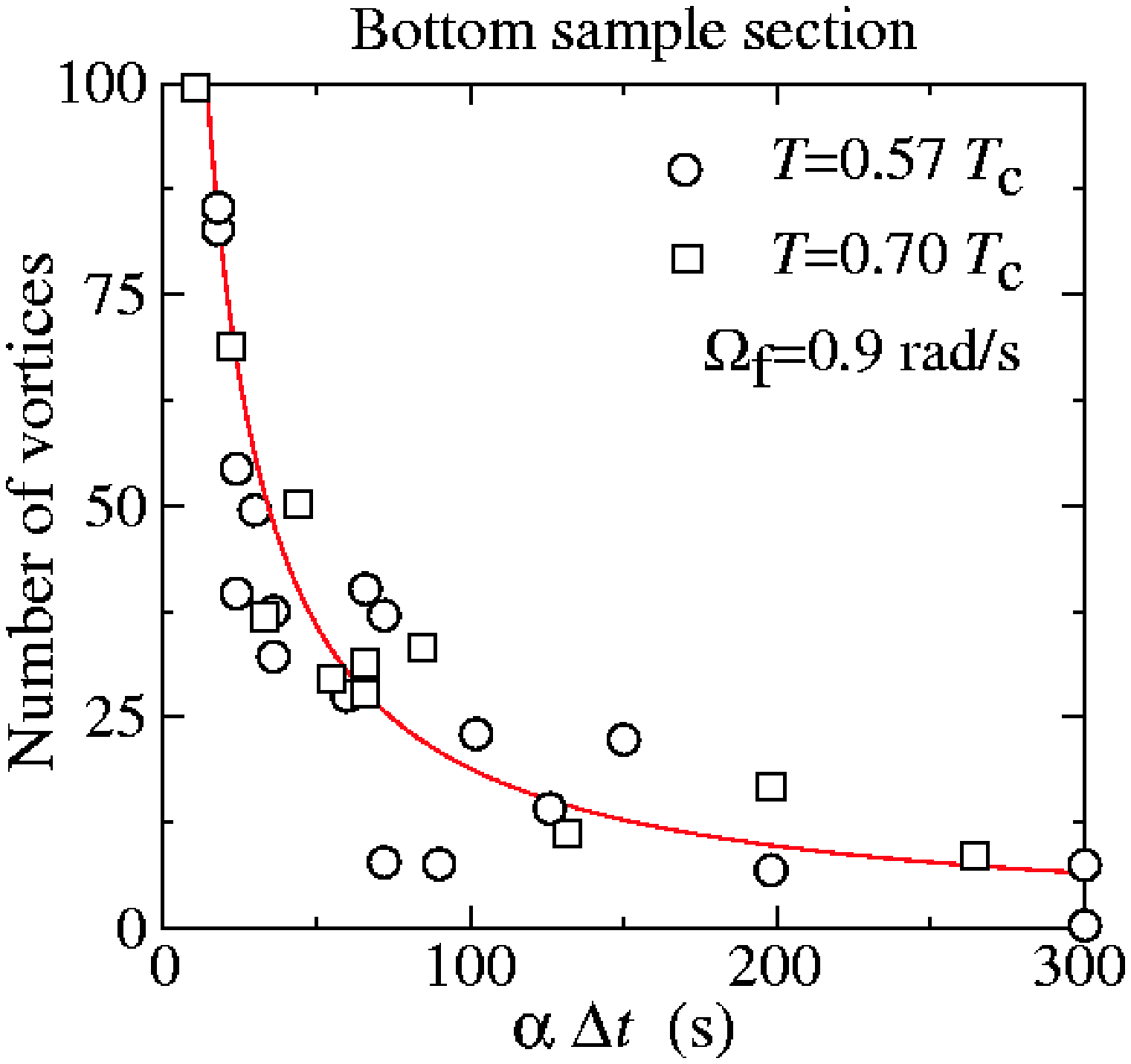}
\caption{Number of remnant $\cal{N}$ for A-phase separated top and
bottom sample sections, as a function of the waiting period
$\Delta t$ at zero rotation. The results can be fit with
$\cal{N}$$(\Delta t) \approx 2 \cdot 10^3 /(\alpha \, \Delta t + b
)$ [where $\Delta t$ is in sec and $b \approx 7\,$s allows for the
initial remnants after deceleration to zero]. The data are for
$\Omega_{\rm f}=0.9\,$rad/s at $0.57\,T_{\rm c}$ with $\alpha =
0.60$ and at $0.7\,T_{\rm c}$ with $\alpha = 1.1$.}
\label{RemnantVorWaitTime}
\end{center}
\vspace{-8mm}
\end{figure}

\textbf{Annihilation time of remnants:} In
Fig.~\ref{RemnantVorNumber} the number of remnants $\cal{N}$ is
measured at fixed $T$ and $\Delta t$ as a function of $\Omega_{\rm
f}$. As expected, the results prove to be independent of the
choice of $\Omega_{\rm f}$. The measurement also shows that
$\cal{N}$ scales with the length of the sample, if $L \gg R$. Thus
the annihilation occurs such that the density of remnants
decreases roughly at the same rate everywhere along the sample.
The configuration before final annihilation is the state of
longest life time where only one remnant is present in each cross
section of the cylinder. In a long cylinder many remnants can
still be stacked after another in configurations like that in
Fig.~\ref{RemnantVortex}.

In Fig.~\ref{RemnantVorWaitTime} $\cal{N}$ has been measured as a
function of the waiting period $\Delta t$. The monotonic
annihilation has been recorded independently for the top and
bottom sample sections at two different temperatures. The smooth
decay, with $\cal{N}$$(\Delta t) \propto (\alpha \, \Delta
t)^{-1}\,$, excludes the existence of large numbers of pinned
vortices in pinning configurations larger than the equivalent of
$r_{\rm o} \sim 3\,\mu$m. The decay of dynamic remnants is thus
governed by $\Delta t$ and $\alpha (T)$ which control their
number, size, and configuration. To reach vortex-free flow at
decreasing temperatures, one has to extend $\Delta t$ to longer
times.

\begin{figure}[t]
\includegraphics[width=0.95\linewidth]{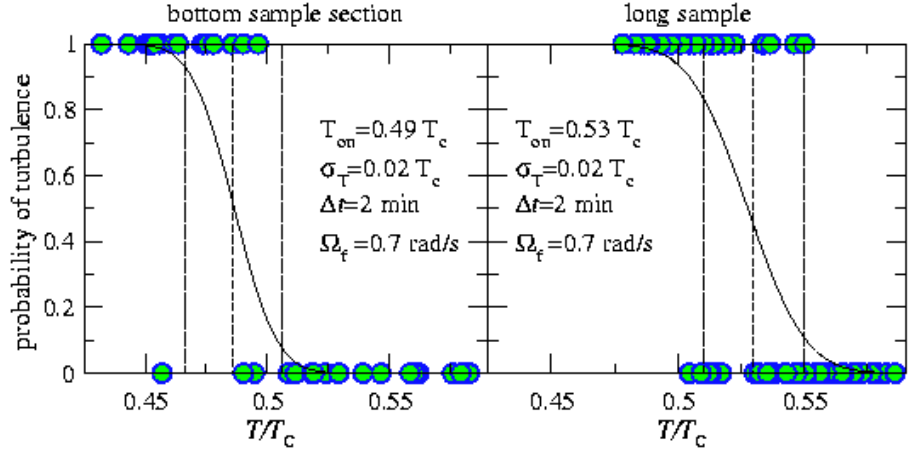}
\vspace{-5mm}
\begin{center}
\caption{Onset temperature $T_{\rm on}$ of turbulence, shown here
for the A-phase separated bottom section (left) and the long
sample with no A phase (right). The probability of turbulence at
$\Omega_{\rm f} = 0.70\,$rad/s is plotted as a function of
temperature with $\sim 40$ data points per panel. The A-phase
separated top section has an onset distribution with the same half
width $\sigma_{\rm T} = 0.02\,T_{\rm c}$, but centered around
$T_{\rm on} = 0.44\, T_{\rm c}$. } \label{RemnantOnset}
\end{center}
\vspace{-12mm}
\end{figure}

\textbf{Onset of turbulence:} In Fig.~\ref{RemnantOnset} the onset
temperature $T_{\rm on}$ to turbulence is measured. Each data
point is obtained at constant temperature, starting from a state
with the remnants after a waiting period of $\Delta t = 2\,$min.
On the vertical axis we plot the probability of finding the
equilibrium vortex state at $\Omega_{\rm f} = 0.7\,$rad/s (with
$N_{\rm eq} \approx 500$ for short sample and 560 for long
sample). If this state is observed then the turbulent burst has
occurred. In the opposite case only a small number of rectilinear
vortices is found ($N \lesssim 35$ in short sample and $\lesssim
55$ in long sample). Such cases where the final number of vortices
is clearly between these values are rare and limited to the onset
regime around $T_{\rm on}$. The $T_{\rm on}$ data fit the normal
distribution with a half width $\sigma_{\rm T} \sim 0.02\,T_{\rm
c}$.

Compared to the top sample section, the bottom section displays
(1) a higher $T_{\rm on}$ and (2) a weaker dependence of $T_{\rm
on}$ on $\Delta t$ and the number of remnants $\cal{N}$. We
attribute these differences to the presence of the
orifice\cite{ExpSetup} on the bottom of the sample cylinder. The
lower values of $T_{\rm on}$ in the top section also mean that the
AB interface does not promote remanence. In
Fig.~\ref{RemnantOnset} the bottom section is compared with the
long sample, so that in both cases the orifice is present. Since
$T_{\rm on}$ is higher for the long cylinder, although the initial
density of remnants is the same in the short and long samples
(Fig.~\ref{RemnantVorNumber}), we conclude that the higher onset
arises because there are twice as many remnants in the long sample
and they spend a longer time evolving to rectilinear lines. Thus
they have a higher probability to become unstable and generate new
vortices during their evolution. At the relatively low dissipation
$\alpha(T) \approx 0.4$ in the onset regime the remnants are
largely insensitive to the direction of rotation after the waiting
period $\Delta t$ and are easily reoriented when flow is
reapplied, even if its direction is opposite from before. In
addition to the number of remnants $\cal{N}$ ({\it i.e.} $\Delta
t$), $T_{\rm on}$ depends also on $\Omega_{\rm f}$. The different
mechanisms are currently under investigation, which influence
$T_{\rm on}$ and cause the transition to turbulence.\cite{ROP}

\textbf{Consequences:} Our results are consistent with the
presence of dynamic remnants -- no support for pinned remnants is
found. In fact, since vortex-free flow down to $0.20\,T_{\rm c}$
is consistently achieved in the present sample cylinder up to
velocities of 3\,mm/s or more by cooling from high temperatures in
rotation, no pinning configurations with $r_{\rm o} \gtrsim
3\,\mu$m exist. It thus appears that sufficiently smooth-walled
containers exist for $^3$He-B in which vortices are not
permanently pinned by surface roughness. 

To achieve vortex-free rotating flow in an open cylinder
non-invasive measurements have to be employed, to avoid internal
probes or any components which enhance pinning because of
non-uniform geometry. Good control of vortex formation allows
studies with new experimental approaches, such as the injection of
seed vortices in applied flow. However, with decreasing
temperature mutual friction dissipation diminishes and vortex
annihilation becomes exceedingly slow. At the lowest temperatures
the last dynamic remnants can only be removed by warming up to
higher temperatures, to speed up annihilation. Using thermal
cycling, relatively high-velocity stable vortex-free flow appears
possible even in the zero temperature limit.

In superfluid $^4$He-II literature it is usual to assume that
remanent vortices originate from surface pinning. Since $\alpha
\lesssim 0.1$ in most of the experimental temperature range of
$^4$He-II, dynamic remanence is more prevalent there than in
$^3$He-B. With careful selection and preparation of the container
walls one can suppress surface roughness to $\lesssim 1\,\mu$m, so
that pinned remnants will remain immobile at $v \lesssim 3\,$mm/s.
In this range of applied flow the present considerations about
dynamic remnants apply equally to $^4$He-II.

\end{document}